 \documentclass[final,authoryear,3p,times]{elsarticle}

\usepackage{graphics}

\usepackage{amssymb}

\journal{Astroparticle Physics}

\begin{document}

\begin{frontmatter}



\title{Studies of active galactic nuclei with CTA}


\author{A. Reimer$^1$ and M. B\"ottcher$^2$}

\address{$^1$Institut f\"ur Theoretische Physik, and\\
Institut f\"ur Astro- und Teilchenphysik,\\
Leopold-Franzens-Universit\"at Innsbruck\\
Technikerstra\ss e 25, A-6020 Innstruck, Austria\\
anita.reimer@uibk.ac.at\\
$^2$Astrophysical Institute\\
Department of Physics and Astronomy\\
Ohio University, Clippinger \# 251B\\
Athens, OH 45701, USA\\
boettchm@ohio.edu}

\begin{abstract}
In this paper, we review the prospects for studies of active
galactic nuclei (AGN) using the envisioned future Cherenkov
Telescope Array (CTA). This review focuses on jetted AGN, which
constitute the vast majority of AGN detected at gamma-ray
energies. Future progress will be driven by the planned lower 
energy threshold for very high energy (VHE) gamma-ray detections to $\sim 10$~GeV 
and improved flux sensitivity compared to current-generation
Cherenkov Telescope facilities. We argue that CTA will enable 
substantial progress on gamma-ray population studies by 
deepening existing surveys both through increased flux sensitivity
and by improving the chances of detecting a larger number of
low-frequency peaked blazars because of the lower energy threshold.
More detailed studies of the VHE gamma-ray spectral shape and
variability might furthermore yield insight into unsolved questions
concerning jet formation and composition, the acceleration of 
particles within relativistic jets, and the microphysics of the
radiation mechanisms leading to the observable high-energy emission.
The broad energy range covered by CTA includes energies 
where gamma-rays are unaffected from absorption while propagating in the 
extragalactic background light (EBL), and extends to an energy regime where
VHE spectra are strongly distorted. This will help to reduce systematic effects
in the spectra from different instruments, leading to a more 
reliable EBL determination, and hence will make it possible to 
constrain blazar models up to the highest energies with less ambiguity.

\end{abstract}

\begin{keyword}
Active galactic nuclei -- Gamma-rays -- Jets


\end{keyword}

\end{frontmatter}


\section{Introduction}
\label{Intro}

\begin{figure}
\centering
\includegraphics[width=0.5\textwidth]{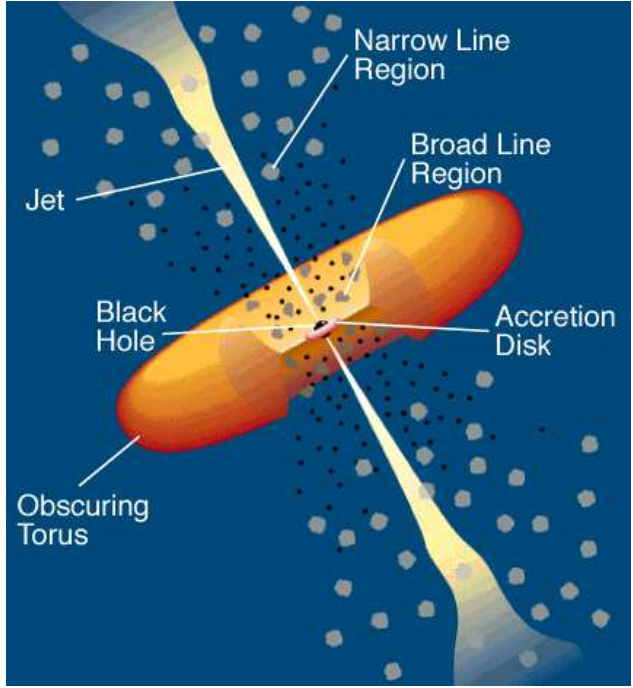}
\caption{Sketch illustrating the constituents and geometry of a
radio-loud AGN \citep[from][]{up95}.}
\label{upsketch}
\end{figure}

Active galactic nuclei (AGN) are extragalactic sources of enhanced activity 
that are powered by the release of gravitational energy from a supermassive 
central black hole. Energy linked to the black hole spin \citep[e.g.,][]{bz77} 
or rotating accretion disks \citep[e.g.,][]{bp82} may be instrumental 
for forming prominent jets which transport material from 
the innermost region of the AGN to kpc-, sometimes even Mpc-scale distances 
with relativistic speed. Such jets are usually identified through the detection 
of bright non-thermal radio emission as observed in radio-loud AGN.  Only a 
small percentage ($\sim 10$~\%) of all AGN are known to be radio-loud\footnote{Radio-loud 
AGN are conventionally characterized with
a radio-to-optical flux ratio $S_{\rm{5GHz}}/S_{\rm{440nm}}>10$.}.
In the vicinity of the central region of an AGN matter is accreted from a 
disk onto the black hole, line-emitting clouds of material (the so-called 
broad-line region, BLR, and narrow line region, NLR) swirl at pc to kpc 
distances from the central engine, dusty material surrounding the accretion 
disk may imprint thermal signatures in the infrared part of the AGN spectrum, and 
the prominent jets of material in case of radio-loud AGN dominate the non-thermal radiative power in such systems (see Fig.~\ref{upsketch}).

The radiation from material which moves relativistically with speed 
$\beta_{\Gamma} c$ (with $\Gamma = 1/\sqrt{1 - \beta_{\Gamma}^2}$ being
the bulk Lorentz factor) along the jet axis is beamed into an angle 
$\sim 1/\Gamma$ around the direction of propagation.
Because of this beaming effect mostly those AGN whose jet axes are close to 
alignment with our line of sight (i.e., blazars) are favourably detected as
sources of high-energy (gamma-ray) emission. However, also some mis-aligned 
AGN (i.e., radio galaxies) can be detected, if they are sufficiently nearby.
Blazars therefore offer an excellent opportunity to
study jet physics of massive black hole systems, and through population 
studies also their evolution over cosmic time. Because the bolometric 
radiative energy output of AGN jets is often dominated by the gamma-ray 
regime\footnote{We note that the apparent dominance of gamma rays in the 
overall blazar budget was recently shown to be affected by selection effects 
\citep{giommi11}.}, the observed peak flux $(\nu F_{\nu})^{\rm pk}$ in this 
band, together with a knowledge of the bulk Lorentz factor $\Gamma$, provides 
a robust lower limit for the overall jet energetics constrained
by the total 
radiative power, $L_{\rm jet} > L_{\rm rad}$, with 
\begin{equation}
L_{\rm rad} \approx {d_{\rm{L}}^2 \over \Gamma^2} \, (\nu F_{\nu})^{\rm pk},
\label{Lrad}
\end{equation}
where $d_{\rm{L}}$ is the luminosity distance. 
Such limit is not only crucial for constraining jet formation scenarios 
and the overall particle and field content of a jet
including its impact for searches for the sources of the ultrahigh 
energy cosmic rays, but also for, e.g., 
investigating the jet's feedback on its environment.
Comparing disk and jet energetics may 
give important clues on the physical connection between disk accretion and 
jetted outflows. Because these jets form in the vicinity of the strong 
gravitational fields of massive, probably rotating, black holes,
studying events occuring close to the central engine may contribute 
to understanding jet formation. Size scales of the emission region of
the order of the Schwarzschild radius are implied by extreme variability 
observed e.g., down to a few minutes time scales at TeV energies 
\citep{aharonian07,albert07}, in a 
few radio-loud AGN, and this might imply a location of the emission region 
very close to the central black hole. 
On the other hand, the observation of systematic variations of the optical 
polarization over several days associated with a gamma-ray flare 
\citep[e.g.,][]{3c279nature}, and distinct gamma-ray flares coinciding 
with the peak polarization of the mm-core \citep{jorstad09} seem to favour rather
pc-scale distances of the emission region relative to the central engine.
This highlights the current debate regarding the location of the emission 
region. Studying the gamma rays from jets within 
the multifrequency context offers a view towards the global structure 
and composition of magnetized relativistic outflows, which provide 
constraints on the dominant radiation mechanisms. Monitoring the 
transition from flaring events to the quiescent phases together with 
the estimates on the overall flaring duty cycles may provide hints on 
the origin of variability.

Gamma rays probe the highest energy particles present in these jets, and 
therefore are relevant for our understanding
of how charged particles are 
accelerated in jet plasmas, e.g., via shocks, and/or turbulence and/or 
magnetic reconnection.
This may also have implications for our understanding 
of the origin of ultrahigh energy cosmic rays.

In this article, we review the prospects of CTA to facilitate progress 
in our understanding of the AGN phenomenon and its related physics 
including the large-scale impact of the associated jets.

\section{CTA and the population of AGN}
\label{Pop}

According to the unification scheme of radio-loud AGN \citep[e.g.,][]{up95},
flat spectrum radio-quasars (FSRQs) and BL~Lac objects (commonly referred 
to as ``blazars'') are sources which are observed under a small viewing 
angle with respect to the jet axis. Those observed at large viewing 
angles are classified as Fanaroff Riley I and II radio galaxies \citep{fr74}.
Hence, they are commonly considered as the parent populations of blazars\footnote{The recently
proposed scenario of \cite{giommi11} considers high-excitation (HERG) and low-excitation (LERG)
radio galaxies as the parent populations of blazars. Nearly all FRIs, however, are LERGs, and
FRIIs are mostly HERGs except for a small FRII LERG population.}.
Jetted AGN are oberved as sources of radiation across the electromagnetic
spectrum, from the radio band up to very-high energy (VHE: $E > 100$~GeV) 
gamma-ray energies. The blazar class is subdivided into observationally weak-lined AGNs, 
identified as BL Lac objects, and strong-lined ones, called flat-spectrum 
radio quasars (FSRQs) \citep{Landt04}. The latter show signatures of a 
bright accretion disk (e.g., a ``blue bump'') and strong emission lines,
whereas the former are lacking such features.
However, the lack of
thermal (accretion disk) and emission line features in the spectra
of objects typically classified as BL~Lac objects may be --- at least in
some cases --- due to the bright non-thermal continuum of jet emission
outshining those features rather than their actual absence \citep{giommi11}.

The spectral energy distributions (SEDs) of jetted AGN consist generally of two broad components
(see, e.g., Fig. \ref{3C66ASEDfit}). The 
low-energy component is commonly attributed to synchrotron radiation 
from relativistic electrons, and possibly positrons, in 
a relativistically moving emission region (``blob'') in the jet. The 
origin of the high-energy emission is still a matter of debate, depending 
strongly on the overall jet composition (see below). Spectrally, blazars 
can be classified according to the frequency of the synchrotron peak
in their broadband SED, independent of the optical emission-line based 
characterization of being a BL~Lac object or a quasar \citep{FermiBlazars}. 
Low-synchrotron-peaked (LSP) blazars have their low-energy 
peak at $\nu_{\rm{s,peak}} < 10^{14}$Hz, intermediate-synchrotron-peaked 
(ISP) blazars at $10^{14} \leq \nu_{s,peak} \leq 10^{15}$Hz, and 
high-synchrotron-peaked (HSP) blazars at $\nu_{\rm{s,peak}} > 10^{15}$Hz. 
Considering the BL~Lac population only, we shall distinguish 
low-frequency-peaked BL Lacs (LBLs), intermediate-frequency-peaked 
BL Lacs (IBLs) and high-frequency-peaked BL Lacs (HBLs), correspondingly.

The census of gamma-ray detected blazars has recently experienced a 
dramatic increase from less than 100 blazars detected by the EGRET 
instrument onboard the {\sl Compton Gamma-Ray Observatory} and 
Cherenkov telescopes to nearly $10^3$ high-latitude objects 
detected by the Large Area Telescope (LAT) on board the {\it Fermi
Gamma-Ray Space Telescope} \citep{2LAC}, that have been associated with AGN. 
Only a few non-blazar AGN are detected at gamma-ray energies 
to date: nearly a dozen radio galaxies, a few Narrow-Line 
Seyfert 1 (NLS1) galaxies and a few unusual sources that 
escaped a convincing identification so far \citep{2LAC}.

Until now, there is no convincing case of a gamma-ray detection of a 
``classical'' radio-quiet Seyfert galaxy. An upper limit for the GeV 
luminosity of hard X-ray selected radio-quiet Seyferts as a class
is currently probing the level of about 1~\% of their bolometric 
luminosity, corresponding to a few $10^{-9}$~ph~cm$^{-2}$~s$^{-1}$ 
in the 0.1 -- 10~GeV band \citep{LATSeyferts}. With CTA at its 
predicted sensitivity at low energies it will be possible to extend 
this energy range to several tens to hundreds of GeV at a comparable
energy flux level. This would probe whether there exists a smooth 
extension of radio-loud low-luminosity AGN towards the Seyfert population
\footnote{We note that the conventional definition of a radio-loud AGN may
allow accretion-dominated HSPs and/or AGN with a comparably weak 
non-thermal emission component be possibly classified as radio-quiet.} 
both spectrally as well as regarding their luminosity function, or
whether those are distinct source classes. 

\begin{figure}
\centering
\includegraphics[width=0.5\textwidth]{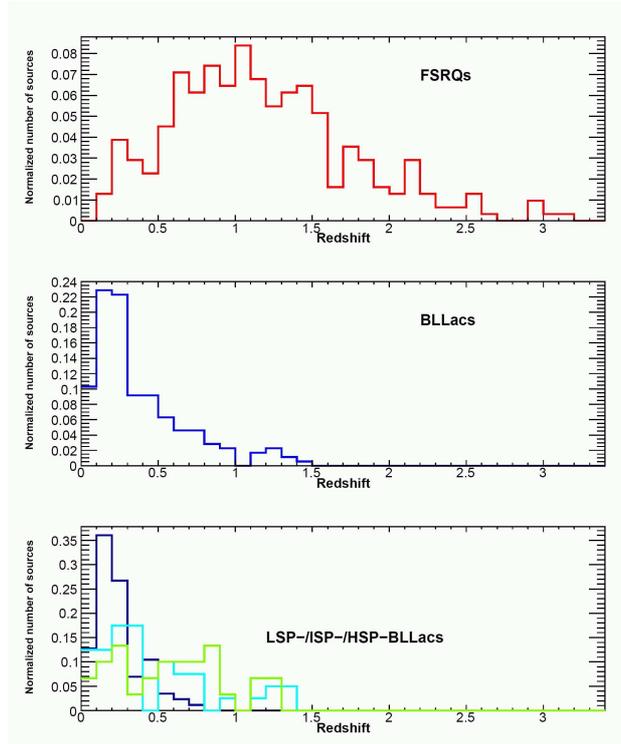}
\caption{Redshift distribution of {\it Fermi}-LAT detected blazars
\citep[from][]{2LAC}.}
\label{zdist}
\end{figure}

The low number ratio of FR~II to FR~I radio galaxies detected at 
gamma-ray energies to date is surprising in the framework of 
the unification scheme. Though the doubling of the {\it Fermi} survey 
time from one to two years has increased the overall number of 
detected gamma-ray emitting AGN by $\sim 50$~\%, the relative 
number of LAT FSRQs to LAT BL~Lacs has decreased from $\sim 0.9$ to 
$\sim 0.8$\footnote{This discrepancy may be even larger due to the many
non-associated BL Lacs because of either the poor signal-to-noise 
ratio in the lines measurements or incomplete cataloguing at the southern 
hemisphere.}. 
The number of gamma-ray detected misaligned AGN has not changed 
significantly, and neither has the FR~II to FR~I number ratio \citep{2LAC}.
Three of the four radio galaxies detected at VHE gamma-ray energies 
(M87, Cen~A, NGC~1275) belong to the FR~I class \citep[the fourth
one,
IC~310, is of unknown class, possibly a head-tail radio galaxy:][]{IC310}. 
A deeper survey of
FR~I and FR~II radio galaxies with CTA at energies $> 10$~GeV 
is
expected to increase the sample of VHE gamma-ray emitting radio
galaxies and may lead to the detection of a few FR~II radio galaxies. 
This might facilitate the determination of the ratio of VHE gamma-ray 
emitting FR~II to FR~I sources over a broader energy range. 
Such studies might reveal whether the 
less prominent gamma-ray emission from FR~IIs is indicative of less 
efficient particle acceleration in their jets, a difference in jet structure 
(e.g., \cite{chiaberge00}) and/or beaming pattern between FR~Is/BL~Lacs and 
FR~IIs/FSRQs (e.g., \cite{dermer95}),  whether possibly 
$\gamma\gamma$ absorption in the dense nuclear radiation fields of 
these generally more powerful sources 
\citep{reimer07,liu08,sb08,ps10,rb10,rb11} plays a role in 
suppressing observable gamma-ray emission, or whether FRII/HERGs 
are intrinsically less numerous
as a consequence of being located at the high-luminosity end of an overall
radio galaxy luminosity function 
\citep{giommi11}.

\begin{figure}
\centering
\includegraphics[width=0.8\textwidth]{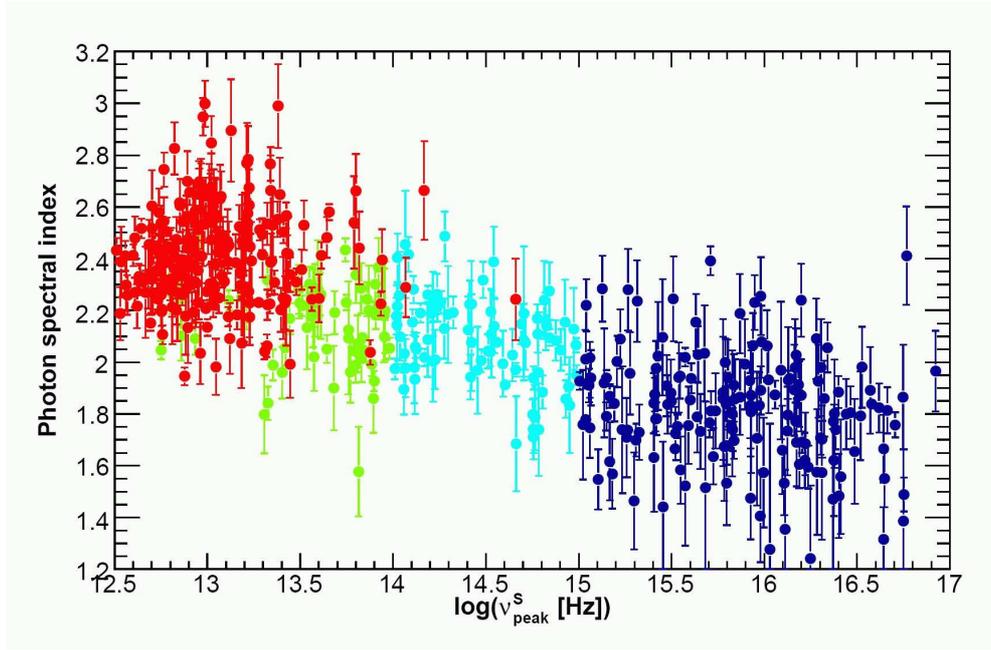}
\caption{Synchrotron peak frequency vs. {\it Fermi}-LAT spectral
index for {\it Fermi} detected blazars \citep[from][]{2LAC}.
Red = FSRQs; green = LBLs; light blue = IBLs; dark blue = HBLs.}
\label{nus_Gamma}
\end{figure}

\begin{figure}
\centering
\includegraphics[width=0.8\textwidth]{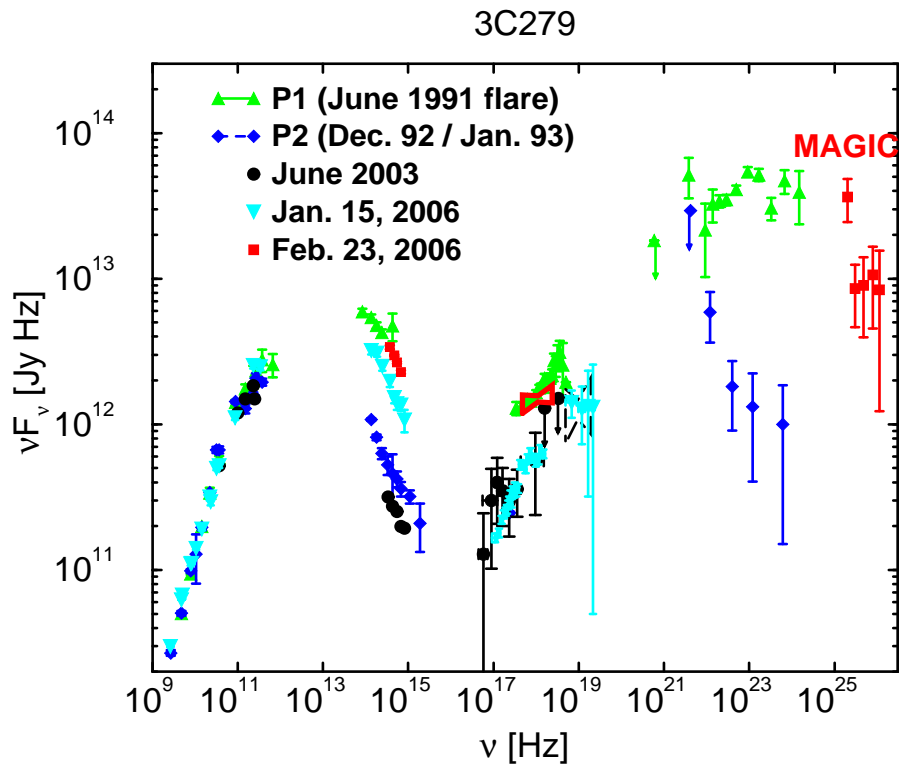}
\caption{Broadband spectral energy distribution of FSRQ 3C~279 during the June 1991 and 
February 2006 flaring state in comparison to 1992/1993 and 2003 observations where the 
source was in a quiescent state
\citep{brm09}.}
\label{3C279}
\end{figure}

Among the LAT-detected BL~Lacs the high-synchrotron peaked sources 
(HSPs) are the largest subclass, which is also the AGN subclass that 
is mostly detected in the VHE-regime by current Atmospheric Cherenkov 
Telescope (ACT) instruments. The (nearly permanent) survey observation 
mode of {\it Fermi}-LAT has triggered many follow-up observations of 
selected flaring AGN also with H.E.S.S., MAGIC and VERITAS.
Until a 
few years ago, almost all AGNs detected by ground-based Cherenkov 
telescopes were HSPs, primarily because of their harder GeV gamma-ray 
spectra (see Fig.~\ref{nus_Gamma}), indicating higher gamma-ray 
peak frequencies than other blazar subclasses. However, due to 
their permanently improving flux sensitivity and decreasing 
threshold energies, more than 40 blazars of all subtypes (FSRQs, 
all types of BL Lac objects: LBLs, IBLs, HBLs) have meanwhile been 
detected in VHE gamma-rays, covering the redshift range 0.03 to 
at least 0.536, thereby nearly doubling the census
of VHE blazars 
during the past couple of years\footnote{see {\tt http://www.mpp.mpg.de/~rwagner/sources/} 
or {\tt http://tevcat.uchicago.edu/}}. This may indicate that with 
CTA it will be possible to significantly expand the population 
of low-frequency peaked VHE AGN, both FSRQs and BL Lac 
objects, in addition to the HSP population. 
A systematic unbiased study may reveal then the required environment
and jet properties that allows particles and photons to reach high energies.
In particular, future observation of a larger number of BL Lac -- FSRQ transition objects
in the VHE gamma-ray band, accompanied by multifrequency coverage, may 
provide more insight in this regard. So far only few (e.g., 3C~279 whose thermal
components may be overwhelmed by a strong non-thermal flux in a bright state \citep{pian99};
or the BL Lac prototype, BL Lacertae, which shows occasionally broad lines 
\citep[e.g.,][]{capetti10})
of such objects have been detected at VHEs when they reached an elevated 
flaring state \citep{albert08,brm09,albert07b}. 
E.g., while monitoring 3C~279 in 2006 in a dedicated multifrequency campaign 
\citep[][see Fig.~\ref{3C279}]{b07} this source transitioned to an overall 
high state observed in the optical as well as at X-rays. During this state, in February 2006,
a VHE signal from 3C~279 was detected by the MAGIC telescope \citep{albert08}.
The observations of a larger sample of such objects with CTA might uncover a 
particular spectral and/or variability pattern with possible relations 
to other frequency bands that may help to finally reveal
the conditions in the jet that allows charged particles to reach extreme energies.
A similar consideration can be applied to LBL -- HBL transition AGN. Broadband studies of such
objects, including the important VHE regime, may shed light on the physical origin of such
behaviour, and will help to determine to what extent LBLs and HBLs are fundamentally 
different AGNs. 

Both the expected 
increased sensitivity of CTA and extension of the available energy 
range towards tens of GeV will also facilitate studies of the AGN 
population at VHE gamma-rays to very large redshifts. The so far highest-redshift
source detected at VHEs is 3C~279 at $z=0.537$.
The redshift range of AGNs covered by {\it Fermi}-LAT extends to
$z<3.1$ (unchanged from the first to the second year of {\it Fermi} exposure; see Fig.~\ref{zdist}), 
above $\sim 20$~GeV it is $z<3$ \citep{2LAC},  while FSRQs are known to exist up 
to $z\sim5.5$ (e.g., Q~0906+6930: \cite{romani06}). 

With CTA, a new quality of the study of AGN evolution over cosmic time
will be possible. The VHE range is important as it provides an undiluted 
view on the pure jet. Proposed cosmological evolution scenarios \citep{bd02,cd02} 
consider a gradual depletion of the circum-nuclear matter and radiation fields 
over cosmic time thereby turning highly-accreting
into pure non-thermal jet systems. This would suggest a 
transition from external-Compton to synchrotron-self-Compton dominated 
high-energy emission in the framework of leptonic emission scenarios 
or from photo-pion dominated to proton-synchrotron dominated high-energy 
emission in the framework of hadronic emission scenarios (see 
\S \ref{Physics}). If this scenario is correct, a systematic study 
of the sub-GeV -- TeV spectra of the various subclasses of blazars 
should therefore reveal a gradual transition from multi-component
gamma-ray emission in accretion-dominated blazars to featureless single-component
gamma-ray emission in pure jet sources.

For the first time, it will be possible to build large, well-defined, 
statistically complete\footnote{A
completely identified sample of sources where all  
are detectable above its statistical limits is generally 
referred to as a complete sample.} and 
unbiased\footnote{We refer to a sample above its statistical limits as unbiased with respect to 
pre-specified parameters if the process that selects the sample sources
does not favour or disfavour any objects with particular values
of the considered parameters.} 
samples at VHEs which allow us to derive population properties like the 
VHE Log(N)-Log(S) distribution for the various types of AGN, the 
luminosity function at VHE gamma rays and compared to other 
wavelengths, and to study their cosmological evolution.
This will extend our knowledge on the origin of the extragalactic gamma-ray background
(e.g., \cite{lognlogs}) up to the highest photon energies, and its impact on
the evolution of the intergalactic medium and structure formation \citep{puchwein11}.

As the AGN population detected at VHE gamma rays will penetrate to 
larger redshifts, predominantly the high luminosity tail of this population 
will be detected. In particular, verifying the existence or non-existence 
of a high-luminosity HSP population and its broadband spectral properties 
will be interesting as this would contradict the traditional understanding
of the blazar sequence \citep{fossati98}\footnote{However, selection 
effects and other
sample biases may impact the physical existence and 
significance of this proposed sequence \citep[for a review, see][]{padovani07}}.
According to this picture, a sequence of blazar subclasses has been 
proposed to be linked to their bolometric luminosity, black hole mass, 
accretion disk luminosity and accretion mode \citep[e.g.,][]{ghisellini11}. 
It has been suggested that FSRQ activity is powered by accretion at a high 
Eddington ratio ($\dot M / \dot M_{\rm Edd} \gtrsim 10^{-2}$), which might 
be related to dense circum-nuclear environments. The corresponding 
dense circum-nuclear radiation fields are expected to leave
their imprints in two-component gamma-ray spectra as well 
as potentially $\gamma\gamma$ absorption features, if
the 
gamma-ray production zone is located within the broad-line
region of the AGN \citep{ps10}. At the same time, efficient radiative
cooling of relativistic particles in these dense radiation fields
might then impede their acceleration to very high energies, resulting
in SED peaks at low frequencies. On the other hand, BL~Lac objects
are suspected to be powered by radiatively inefficient accretion
at low rates ($\dot M / \dot M_{\rm Edd} \ll 10^{-2}$), possibly
--- at least in part --- due to larger black-hole masses (and
hence larger $\dot M_{\rm Edd}$). If the jet power correlates
positively with the accretion rate \citep[e.g.,][]{rs91}, this
implies a lower power in the jets produced in these objects,
compared to FSRQs. At the same time, the circum-nuclear radiation 
fields are expected to be very dilute, with only minor impact on 
the formation of the high-energy (gamma-ray) emission. 
The search for high-luminosity (high-redshift) BL~Lac objects
with high synchrotron and gamma-ray peak frequencies with
CTA, in combination with on-going monitoring by {\it Fermi}-LAT
promises progress in verifying the existence of and understanding 
the origin of the blazar sequence, or whether the peak energy is 
intrinsically unrelated to the blazar luminosity \citep{giommi11}. 
The redshift of these objects, if lacking as argued by \citet{giommi11},
could be constrained using UV-to-NIR photometry \citep{rau11}, or limits inferred
from the shape of the de-absorbed spectrum if the extragalactic background 
light (EBL) and its evolution were known \citep[e.g.][]{1553,prandini11}.

\begin{figure}
\centering
\includegraphics[width=0.7\textwidth]{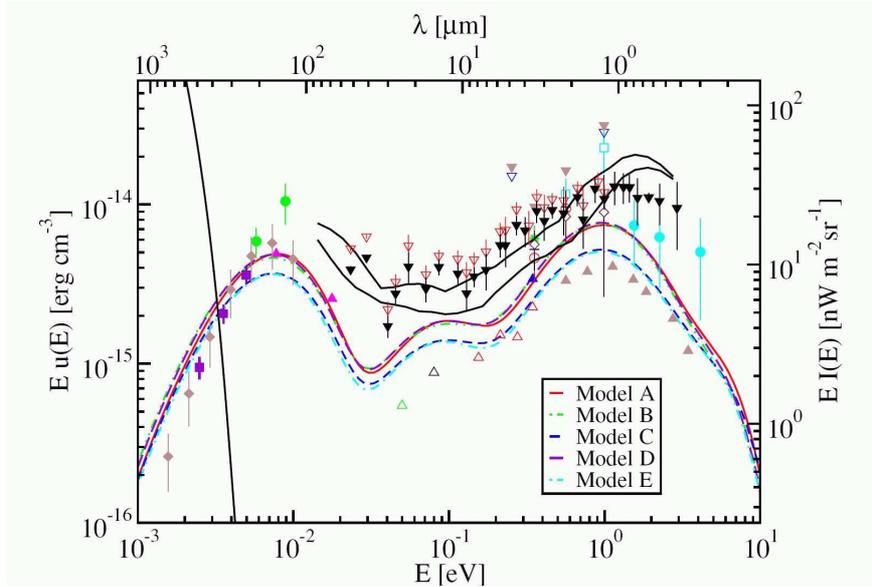}
\caption{Spectral energy distribution of the Extragalactic Background
Light. From \cite{finke10}.}
\label{EBLfig}
\end{figure}

\section{\label{EBL}The extragalactic background light and blazar spectra}

\begin{figure}
\centering
\includegraphics[width=0.7\textwidth]{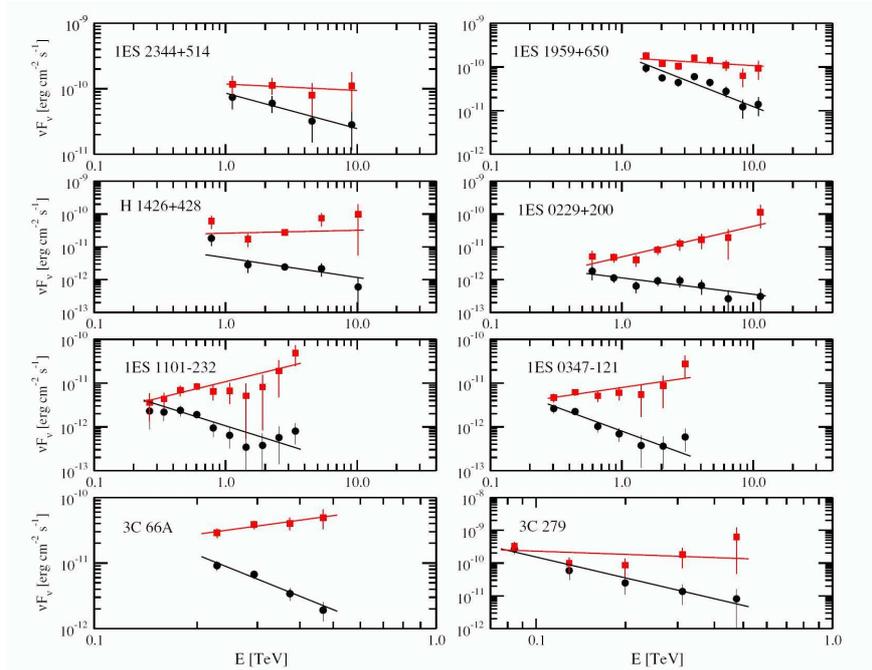}
\caption{VHE gamma-ray spectra of eight blazars at various redshifts,
corrected for EBL absorption using the model of \cite{finke10}.}
\label{EBLabs}
\end{figure}

VHE gamma-rays from sources at cosmological distances will be
attenuated through $\gamma\gamma$ absorption on the extragalactic
background light \citep[EBL; e.g.,][]{dk05,sms06,franceschini08,gilmore09,finke10}. 
The SED of the EBL has two maxima: one at $\sim 1 \,\mu$m due to
star light from cool stars, and one at $\sim 100 \, \mu$m due to
cool dust (see Fig.~\ref{EBLfig}). A direct measurement of this 
background is extremely difficult because of bright foreground 
emissions (both within our solar system and our Galaxy). The recent 
measurements of unexpectedly hard VHE gamma-ray spectra from blazars 
at relatively high redshifts (see, e.g., Fig.~\ref{EBLabs}) has led 
to the conclusion that the intensity of the EBL must be near the lower 
limit set by direct galaxy counts \citep[e.g.,][]{aharonian06,abdo10},
or that the gamma-ray signal might be contaminated by ultra-high energy cosmic 
ray-induced photons \citep[e.g.,][]{essey10}. A more exotic alternative
explanation that has been proposed is that VHE $\gamma$-ray photons may
be converted to axion-like particles when interacting with magnetic 
fields either in the vicinity of the blazar or in intergalactic space. 
Those particles would be able to travel to Earth unaffected by the EBL, 
and may be re-converted to $\gamma$-rays in interactions with Galactic 
magnetic fields \citep{deangelis07,simet08}. Even assuming that EBL
absorption is not circumvented, details of the spectral shape and, 
in particular, the cosmological evolution of the EBL are still 
uncertain.

Indirectly, the EBL and its cosmological evolution can be studied by 
analyzing simultaneous broadband SEDs of VHE gamma-ray blazars at 
various known redshifts. In particular, simultaneous {\it Fermi}-LAT and
ground-based VHE gamma-ray spectra are crucial for such an analysis.
However, this requires an a priori knowledge of the source-intrinsic 
SED throughout the 
GeV -- TeV energy range. The uncertainties and ambiguities in blazar 
jet models (see \S \ref{Physics}) currently preclude definite
conclusions about the EBL based on blazar SED modeling alone. An
observational challenge in such studies lies in the often vastly
different integration times over which gamma-ray spectra in the
{\it Fermi}-LAT energy range are measured (typically several weeks),
compared to VHE gamma-ray spectra, often extracted from a few hours
of good data from ground-based ACTs. This often leads to mis-matches
in the spectral shapes and flux normalizations, which complicates
or impedes any meaningful theoretical interpretation.

With the reduced energy threshold of CTA, down to $\sim 10$~GeV,
it will be possible to determine the shape of the gamma-ray
spectrum from energies at which EBL absorption is negligible
(typically below a few tens of GeV) out to $> 100$~GeV energies
where the spectrum might be significantly affected by EBL
absorption, depending on redshift. The significant overlap
with {\it Fermi}-LAT could then potentially also 
allow for a more reliable cross-calibration between LAT 
and ground-based ACTs. Given the often very moderate 
variability of the gamma-ray spectral indices of many 
LAT-detected blazars \citep{1LAC}, the cross-calibration with
CTA might then allow for the construction of reliable, truly
simultaneous gamma-ray SEDs through the LAT and VHE gamma-ray
energy ranges. Note that limited correlated flux variability 
between the GeV and TeV energy range of prominent TeV-blazars has 
been observed so far (e.g., during intensive campaigns performed on 
Mkn 501 \citep{mkn501}, Mkn 421 \citep{mkn421}, or PKS 2155-304 
\citep{pks2155}). Simultaneous multiwavelength data sets at lower
wavelengths may then be used to constrain SED models for
a meaningful study of EBL absorption effects at the highest
energies. We caution, however, that the overlap between the 
operations of the LAT and CTA could be extremely limited which 
leads to a correspondingly lower scientific return in 
this regard.

\section{The physics of extragalactic jets}
\label{Physics}

Active galactic nuclei are thought to be systems that are powered 
by the release of gravitational energy.
How, where and in which 
form this energy is released, and especially the physics governing 
to the formation, acceleration and collimation of relativistic jets 
and the conversion of jet power into radiative power is poorly
understood \citep[for a review of the current status of the field,
see, e.g.,][]{jetbook}. The observed links (see \S \ref{Intro}) 
between enhanced emission at high photon energies and changes in 
the polarization properties in the emission region may indicate 
an important impact of the magnetic field topology and strength
on the broadband spectral variability behaviour of jetted AGN
and possibly on the intrinsic acceleration of jet knots \citep[e.g., 
by magnetic driving: ][]{vlahakis04}.
As we will outline below, studies of the 
SEDs and variability of blazars
with CTA, {\it Fermi}-LAT, and 
co-ordinated observations at lower
frequencies will be crucial 
to gain insight into these issues. 

\subsection{Radiative processes in extragalactic jets}
\label{Process}

Depending on the jet's relativistic matter composition two types of emission models 
have emerged during the last decade. {\it Leptonic} models consider 
relativistic electrons and positrons as the dominating emitting relativistic particle
population, while in {\it hadronic}\footnote{So-called 
{\it lepto-hadronic emission models} follow the same physics as 
{\it hadronic emission models}.} emission models the relativistic 
jet material is composed of relativistic protons
and electrons
\citep[for a recent review of 
blazar emission models, see][]{boettcher10,ReimerMuonio11}. In both scenarios, cold 
(i.e., non-relativistic) pairs and/or protons may exist as 
well, allowing charge neutrality to be fulfilled. There is 
meanwhile mounting evidence that jets of powerful AGN have 
to be energetically and dynamically dominated by protons and/or 
ions \citep[see e.g.,][]{cg08}, albeit little is known about 
their spectral distribution (cold, relativistic) and number 
density with respect to the electrons.

In both leptonic and hadronic models, the low-frequency emission 
is produced as synchrotron radiation of relativistic electrons in
magnetic fields in the emission region, which is moving with relativistic
speed corresponding to a bulk Lorentz factor $\Gamma$ along the jet. 
For ease of computation, the magnetic field is typically assumed to
be tangled (i.e., randomly oriented), and the electron distribution
is assumed to be isotropic in the co-moving frame of the emission
region.

In leptonic models, the high-energy emission is produced via Compton
upscattering of soft photons off the same ultra-relativistic electrons
which are producing the synchrotron emission. Both the synchrotron 
photons produced within the jet \citep[the SSC process:][]{mg85,maraschi92,bm96}, 
and external photons (the EC process) can serve as target photons for Compton 
scattering. Possible sources of external seed photons include the accretion 
disk radiation \citep[e.g.,][]{dermer92,ds93}, reprocessed optical -- UV emission
from circumnuclear material \citep[e.g., the BLR:][]{sikora94,dermer97},
infrared emission from warm dust \citep{blaz00}, or synchrotron emission 
from other (faster/slower) regions of the
jet itself \citep{gk03,gt08}.

Relativistic Doppler boosting allows one to choose model 
parameters in a way that the $\gamma\gamma$ absorption opacity 
of the emission region is low throughout most of the high-energy
spectrum (i.e., low compactness). However, at the highest photon energies,
this effect may make a non-negligible contribution to the formation of
the emerging spectrum \citep{aharonian08} and re-process some of 
the radiated power to lower frequencies. The resulting VHE gamma-ray
cut-off or spectral break, and associated MeV -- GeV emission features 
may be revealed by high-resolution, simultaneous {\it Fermi} and CTA 
observations.

\begin{figure}
\centering
\includegraphics[width=0.8\textwidth]{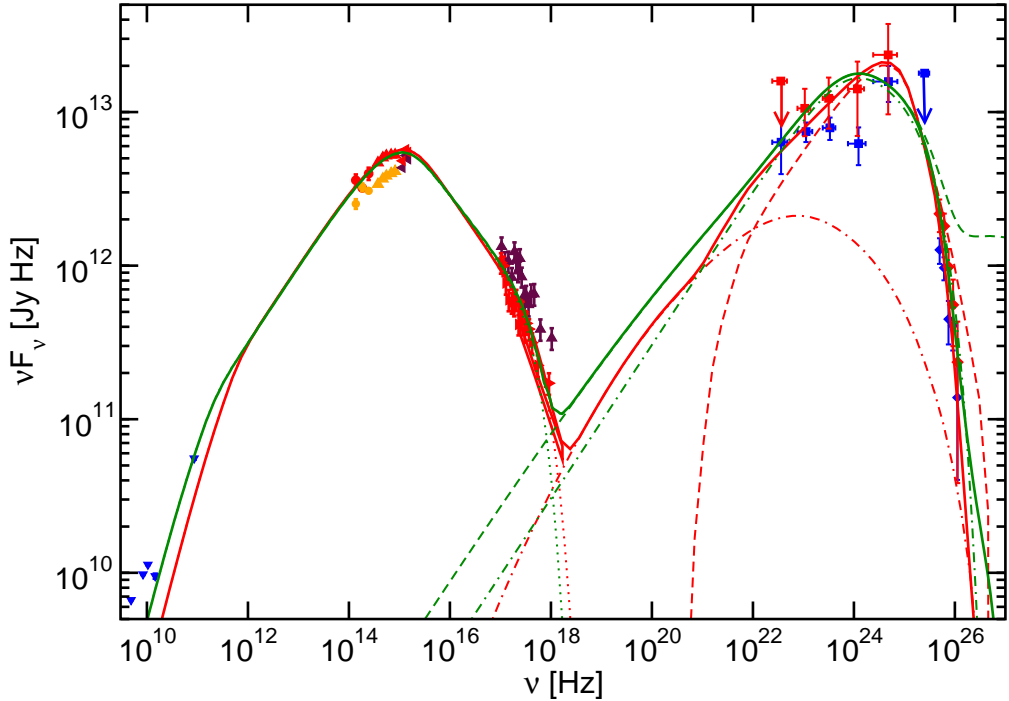}
\caption{Spectral Energy Distribution of the intermediate BL~Lac
Object 3C66A during its bright gamma-ray flare in 2008 October
\citep[from][]{3C66AVERITAS}. Red = leptonic SSC + EC
fit; green = hadronic model fit.}
\label{3C66ASEDfit}
\end{figure}

Hadronic models consider a significant ultra-relativistic proton 
component
in addition to primary ultra-relativistic electrons, to be present 
in the AGN jet. The charged particles interact with magnetic 
and photon fields. In heavy jet models the interaction of protons/ions 
with matter (via e.g., relativistic blast waves \citep{pohl00}, 
star/cloud-jet interaction \citep{Bednarekproc99,Beall99,araudo2010}, 
jet-red giant interaction: \citep{barkov10}) may dominate. 
However, such models \citep[e.g.,][]{reynoso11} do often not predict 
rapid flux variability. Particle-photon interaction 
processes in hadronic models include photomeson production, Bethe-Heitler 
pair production for protons, and inverse Compton scattering of pairs. 
An inevitable by-product of hadronic interactions is the production 
of neutrinos. The target photon fields for such processes include 
internal jet synchrotron photon fields \citep{Mannheim92,Muecke01,Muecke03}, 
and fields external to the jet such as direct accretion disk radiation 
\citep{Bednarek99}, jet or accretion disk radiation reprocessed in the 
BLR \citep{Atoyan03}, the radiation field of a massive star in the 
vicinity of the jet \citep{Bednarek97}  or infrared radiation by warm 
dust \citep[e.g.,][]{dermer12}. The secondary 
particles and photons from interactions of ultra-relativistic hadrons in 
general initiate synchrotron and/or Compton-supported pair cascades
which redistribute the power from very high to lower energies 
\citep[e.g.,][]{Muecke03}. For high magnetic field strengths, any IC 
component is in general strongly suppressed, leaving the 
proton-initiated radiation as the dominating high energy emission 
component.

Figure \ref{3C66ASEDfit} compares a steady-state leptonic (SSC \& EC) fit to a
corresponding hadronic fit of the SED of the IBL 3C66A detected in VHE gamma-rays
by VERITAS in 2008 \citep{3C66Adiscovery,3C66AVERITAS}. Both leptonic
and hadronic models provide excellent fits to the simultaneous SEDs
obtained during the prominent 2008 October gamma-ray flare, with
plausible physical parameters.

Because hadronic interactions convert some protons into 
neutrons\footnote{e.g., in hadronic p$\gamma$-interactions 
30~\% -- 70~\% of the initial protons are converted into 
neutrons \citep{Muecke00}} via charge exchange, collimated 
neutron beams may form \citep{Eichler78,Atoyan03} which
can transport a significant portion of the initial energy to 
large distances from the black hole. When such powerful 
jets interact with the intergalactic medium, large amounts 
of their power and momentum are expected to be deposited 
into the surroundings as huge lobes. The good angular resolution 
of CTA may permit the imaging of such extended emission, and 
will provide valuable information about the total power stored 
in jets, which in turn may constrain jet formation scenarios 
and jet composition.

Because of the suppression of the Compton cross section in
the Klein-Nishina regime\footnote{$\epsilon\gamma \gtrsim 1$, 
where $\epsilon = h \nu / [m_e c^2]$ is the dimenionless photon
energy and gamma the electron Lorentz factor.} and efficient
radiative (synchrotron + Compton) cooling, leptonic models are 
typically hard-pressed to explain hard (energy spectral index 
$\alpha \lesssim 1$, where $F_{\nu} \propto \nu^{-\alpha}$) 
gamma-ray spectra extending to $E \gtrsim 1$~TeV after
correction for $\gamma\gamma$ absorption by the EBL \citep[e.g.,][]{aharonian06}. Detailed spectral
measurements in the GeV -- TeV regime through simultaneous
observations by {\it Fermi}-LAT and CTA are expected to reveal
the signatures of radiative cooling of leptons and/or Klein-Nishina
effects in leptonic models, or of proton-synchrotron emission
and ultra-high-energy induced pair cascades in hadronic models.
These might therefore distinguish between leptonic and hadronic 
models. 

Simultaneous multi-wavelength coverage will be crucial to put 
meaningful constraints on models. In this context, e.g., \cite{brm09} 
have demonstrated that the extension of the gamma-ray emission 
of the FSRQ 3C~279 into the VHE regime \citep{albert08} poses severe 
problems for homogeneous, leptonic one-zone models, and may favor 
hadronic models, or multi-zone models. The lowered energy threshold 
of CTA compared to current ACTs promises the detection of VHE gamma-ray 
emission from a larger number of low-frequency peaked blazars (including
FSRQs), which will allow for similar studies on a larger sample
of LSP blazars. 

\begin{figure}
\centering
\includegraphics[width=0.8\textwidth]{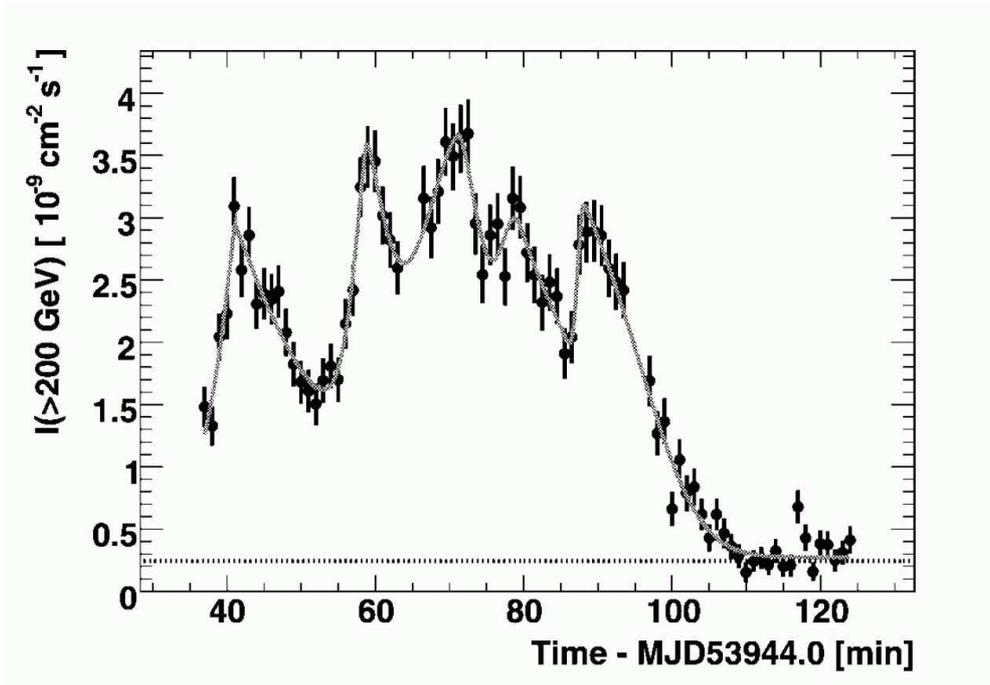}
\caption{Rapid VHE gamma-ray variability of the HBL PKS~2155-304
observed by H.E.S.S. in 2006 \citep{aharonian07}.}
\label{2155_HESS_flare}
\end{figure}

The radiative cooling time scales are generally expected to
be much shorter for leptons than for hadrons. Therefore, measurements
of rapid variability \citep[e.g.,][see also Fig.~\ref{2155_HESS_flare}]{aharonian07,albert07} might 
be an indication for a leptonic origin of (at least parts of) 
the gamma-ray emission from blazars exhibiting variability
on sub-hour time scales. Variability on a few minutes time scale has been 
observed at VHEs from few blazars both of HSP and LSP type (e.g., 
PKS~2155-304 \citep{aharonian07}, Mkn~501 \citep{albert07}, PKS 1222+216 
\citep{aleksic1222}) so far. This implies extremely large
bulk Doppler factors if interpreted within a homogeneous emission 
model, or TeV emitting sub-structures within the jet such as filaments, 
reconnection zones \citep{giannios09}, etc. For example, the spine-sheath 
picture \citep{ghisellini05} of a jet envisions an ultra-fast spine 
surrounded by a slower sheath. If the jet points almost towards the 
observer, radiation from the strongly beamed fast spine dominates the 
observed spectrum, while the radiation from the sheath contributes 
only weakly. In AGN where the jet is more inclined to the sight line
the spine appears as a dim source while the radiation from the slower 
sheath becomes dominant. In order to test this behaviour a larger 
sample of rapidly varying sources, both blazars and radio galaxies,
at VHEs is required. With current technology, only the brightest of 
such sources can be detected, and only in extreme flaring states. The 
increased sensitivity of CTA compared to present-generation 
ACT facilities will allow for the extension of the study of rapid
gamma-ray variability to a large sample of sources and to more 
quiescent states. Variability information in addition to high 
resolution spectra is particularly important for unambiguously 
constraining the parameter space im emission models since in 
many cases (see, e.g., Fig.~\ref{3C66ASEDfit}), pure snap-shot 
SED modeling is unable to distinguish between a leptonic and a 
hadronic origin of the gamma-ray emission.

\subsection{Probing particle acceleration using CTA}
\label{Accel}

Both the SED shape and multi-wavelength variability patterns in 
blazar emission can provide constraints on the mode of particle
acceleration in the jets of AGN. The shape of the high-energy 
end of the particle spectrum --- which will be directly reflected
in the shape of the high-energy end of the gamma-ray emission ---
will provide valuable information about the competition between 
radiative (and possibly adiabatic) losses, escape, and energy 
gain at those energies \citep[e.g.,][]{ps99}. The decreased energy 
threshold and improved sensitivity of CTA over current ACTs will 
enable detailed studies of the shape of the high-energy cut-offs
of blazar spectra (including LSP blazars) and, in particular, 
trace the cutoff in sources not yet detected at VHEs. 

\begin{figure}
\centering
\includegraphics[width=0.6\textwidth]{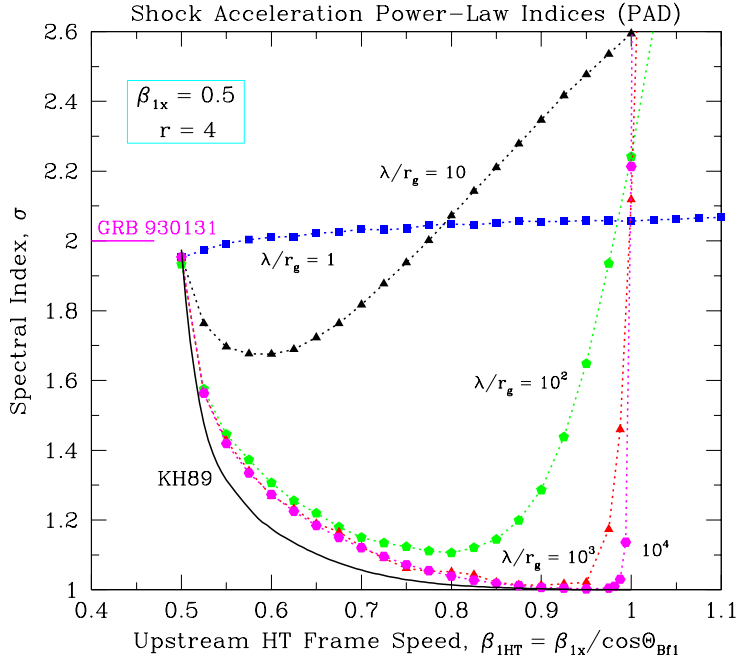}
\caption{Dependence of the relativistic, non-thermal particle
spectral index $\sigma$ on the obliqueness and particle 
mean-free-path $\lambda$ for pitch-angle scattering on
magnetic turbulence. From \cite{baring09}.}
\label{shock_index}
\end{figure}

Different particle acceleration scenarios (e.g., diffusive shock 
acceleration at relativistic shocks, first-order Fermi acceleration, 
perpendicular vs. oblique shocks, diffusive acceleration in shear layers) 
and different magnetic field topologies predict characteristically different 
spectral indices in the resulting 
particle spectra \citep[e.g.,][see also Fig.~\ref{shock_index}]{ob02,so02,ed04,stecker07}. 
These will be directly reflected
in the spectral indices of the non-thermal synchrotron and gamma-ray
emission of blazars. E.g., some HBLs at low fluxes possess very hard 
photon spectral indices (see Fig.~\ref{nus_Gamma}) in the LAT energy range,
implying hard particle spectra of the accelerated particle population. 
CTA might probe the required acceleration conditions in a systematic way. 
Simultaneous multiwavelength observations, including
at the highest energies, will be helpful to probe potential mis-matches
between the low-energy (synchrotron) and high-energy (gamma-ray)
SEDs. In leptonic models, such spectral-index mis-matches typically
require multi-component gamma-ray emission scenarios, if they can
be re-conciled with these models at all. In hadronic models, they might
be explained through different acceleration modes (and hence, different
particle spectral indices) for electrons and protons. 

In addition to simultaneous snap-shot SEDs, spectral variability 
can provide crucial insight into the particle acceleration and cooling
mechanisms in AGN jets \citep[e.g.,][]{krm98,cg99,lk00,bc02}. Detailed 
measurements of spectral variability have so far been restricted to 
lower-energy observations \citep[e.g., X-rays:][]{takahashi96}, or to the brightest
gamma-ray AGN only \citep[e.g., 3C~454.3 at LAT-energies:][]{3C454}. The
improved sensitivity of CTA in the $> 100$~GeV regime might enable
the study of precision spectral variability and persistent long-term variability 
patterns in this energy range for a large sample of sources.
In particular, this will provide a probe of the dynamics of the 
highest-energy particles in LSP blazars in which the high-energy 
end of the synchrotron component is often not observationally
accessible because it is (a) located in the UV/ soft X-ray regime,
which is notoriously difficult to observe, and (b) overlapping
with (and often overwhelmed by) the low-energy end of the high-energy
emission.

\section{\label{Conclude}Concluding remarks}

This surely incomplete list of topics discussed above reveals the 
potential of CTA for significant progress in the field of
AGN research. Improvements in sensitivity and energy coverage will
allow for the study of a much larger population of AGN, although 
we caution that the here important GeV energy range as is currently 
provided by the {\it Fermi}-LAT instrument may be available at the time of 
CTA operations only to an extremely limited extent. This will
enable to tackle a large range of topics from population
studies and questions of cosmological evolution of AGN via studies
of the formation and composition of extragalactic jets and the
microphysics of the production of high energy emission in relativistic
jets, to studies of the Extragalactic Background Light, which will
shed light on the broader issues of cosmological galaxy evolution
and structure formation. 
Most exciting, as CTA will enlarge the dynamical flux range and explore
the high-redshift universe at VHEs, unexpected, possibly surprising, phenomena
may challenge current theoretical concepts, and trigger to deepen our understanding of the
extragalactic sky.
This review might provide some 
insight into possible ways that observations by CTA --- coordinated 
with simultaneous observations at other wavelengths --- might lead
to progress in the study of some of the most pressing questions 
of the VHE sky. 

\vspace*{1cm}

{\bf{Acknowledgements}}

We like to thank Chuck Dermer, Benoit Lott, Marco Ajello and Paolo Giommi for providing
excellent comments on this work which improved this manuscript.
MB acknowledges support from NASA through Astrophysics Theory
Program grant NNX10AC79G and Fermi Guest Investigator Grants
NNX10AO49G and NNX11AO20G. 
AR acknowledges support by Marie Curie IRG grant 248037
within the FP7 Program.


\bibliographystyle{model5-names}
\bibliography{<your-bib-database>}



\end{document}